# Molten *o*-H₃PO₄ – A New Electrolyte for the Anodic Synthesis of Self-Organized Oxide Structures: WO₃ Nanochannel Layers and Others

M. Altomare,[§] O. Pfoch,[§] A. Tighineanu,[§] R. Kirchgeorg,[§] K. Lee,[§] E. Selli,[‖] and P. Schmuki*,[§,†]

[§]Department of Materials Science, Institute for Surface Science and Corrosion WW4-LKO, Friedrich-Alexander University, Martensstraße 7, D-91058 Erlangen, Germany

[†]Chemistry Department, Faculty of Sciences, King Abdulaziz University, 80203 Jeddah, Saudi Arabia Kingdom

[‖]Department of Chemistry, University of Milano, Via C. Golgi 19, I-20133 Milan, Italy

*Corresponding Author. Email: schmuki@ww.uni-erlangen.de

**A B S T R A C T**

We introduce the use of pure molten ortho-phosphoric acid ($o$-$H_3PO_4$) as an electrolyte for self-organizing electrochemistry. This electrolyte allows for the formation of self-organized oxide architectures (one-dimensional nanotubes, nanochannels, nanopores) on metals such as tungsten that up to now were regarded as very difficult to grow self-ordered anodic oxide structures. In this work, we show particularly the fabrication of thick, vertically aligned tungsten oxide nanochannel layers, with pore diameter of *ca.* 10 nm, and illustrate their potential use in some typical applications.

Ever since the work of Masuda *et al.* – that showed that a simple but optimized electrochemical anodization approach can be used to grow, in an entirely self-organized manner, vertically aligned one dimensional (1D) structures[1] – electrochemical anodizing has found wide interest for the controlled and defined growth of nanopores and nanotubular oxide arrays.

While Masuda and follow-up work demonstrated that highly ordered porous alumina can be grown on Al in various aqueous acidic solutions, other reports on anodization found fluoride containing solvents (aqueous or organic) capable of forming highly ordered oxide nanotube structures this first on titanium,[2] then on a wide range of metal substrates.[3,4] In parallel, Melody and Habazaki found that hot glycerol electrolytes[5,6] can provide another platform to establish self-organizing electrochemical conditions (leading to ordered oxide nanochannels on various metals).

In the present work we report on a novel class of a fluoride-free electrolyte, based entirely on molten pure phosphoric acid, which enables the formation of highly ordered, high aspect ratio oxide structures on some elements where self-organizing anodization is considered most challenging. Examples are 1D oxide structures formed from W metal, as well as a full range of other oxide/metal systems such as Nb or Al, as illustrated in Fig. 1.

In the present work we focus particularly on $WO_3$, as the synthesis of such defined $WO_3$ channel structures could not be achieved using other common techniques, such as hydrothermal processes,[7,8] spray pyrolysis,[9] sputtering[10] and thermal/e-beam evaporation.[11,12]

We first explored anodizing in molten *o*-$H_3PO_4$ by an extensive screening of various parameters (provided in the ESI). We observed that namely voltage, anodization time, temperature, and water content were of high importance (the main results are summarized in Table S1 and S2). As a result we found the use of pure (nominally water free, *i.e.*, no added water) molten ortho-phosphoric acid at 100°C, in a range of potential of 1-20 V, to be a most suitable range of conditions for achieving self-organizing anodization.

Under these conditions, the diameter of the anodic films can be controlled by varying applied voltage, and the channel-layer thickness by the anodization time. This is illustrated in Fig. 2 and 3, which show an optimized anodization approach at 5 V that leads to $WO_3$ nanochannels with a mean inner diameter of 10 nm (see also Fig. S1).

Both diameter and length of the nanochannels linearly increase by raising the anodizing voltage from 1 V up to 20 V. Also, for experiments up to 8 h, we found that the layer thickness linearly increased over the anodization time with an average growth rate of *ca.* 0.25 µm h$^{-1}$ (Fig. 3(b), Inset). Most importantly, these films show over the entire parameter range a morphology consisting of straight nanochannels with top open pores. For longer experiments, the thickening of the $WO_3$ layers follows a parabolic trend with anodization time, as etching of the outermost part of the film becomes noticeable (Fig. 3(b)).[13]

The electrolyte temperature was found important to establish an ideal equilibrium between field-assisted passivation and oxide dissolution (that is a prerequisite for self-ordering anodization[14–16]) – in the case of W an *optimum* is obtained at around 100°C. At lower temperatures (60 or 80°C – Fig. S2) only a compact oxide or relatively thin porous layers formed, respectively, while *ca.* 0.5 µm-long nanochannels could be obtained at 120°C (these showed a considerable etching at the outermost part of the anodic film due to the faster oxide dissolution at elevated temperatures).

In regard of the anodizing voltage, we found an applied potential range of 2.5-10 V to be ideal to form thick and ordered layers. The corresponding current density (J) *vs.* time profiles (Fig. S3) indicated that steady state current density values in the range of 0.1-0.6 mA cm$^{-2}$ are required to establish a controlled oxide growth. For lower anodization voltages (1V), only some tens of nm-thick porous layers were formed (with a current density of only a few µA cm$^{-2}$). Potentials of 15 V or higher led to significantly less ordered nanochannel layers (Fig. 2 – in agreement with too high current density values in the range of tens of mA cm$^{-2}$, as shown in Fig. S3).

Another key factor is the water content, this because although nominally water-free molten $o$-H$_3$PO$_4$ is used, the key oxidant is remnant of water. Noteworthy, pure $o$-H$_3$PO$_4$ is sufficiently hygroscopic to maintain a certain water level in the electrolyte. In fact, Karl Fischer analysis (see the ESI for details) demonstrated that the water content of (nominally) pure $o$-H$_3$PO$_4$, molten and heated up to 100°C, is of approximately 1.1% (see Fig. S4), *i.e.*, limited water content in the electrolyte is essential to reach a controlled growth of thick and ordered porous films.

To illustrate the importance of the water concentration we performed anodization experiments (under otherwise optimized conditions) during which a relatively little amount of water was added to the electrolyte (to reach a nominal water content of *ca.* 0.5vol%), and by growing anodic films in $o$-H$_3$PO$_4$-based electrolytes with different initial (nominal) water contents.

As shown in Fig. 4, while water additions up to 2 vol% did not affect the film structure, larger amounts of water (*e.g.*, 10 vol%) led to a strong increase of the etching rate (higher solubility of the anodic oxide) and consequently to a disordered porous film. Water effects are also apparent from the J-time profile in Fig. 4 where a sharp positive spike is observed at the time of water addition. This reflects a sudden acceleration of oxide dissolution. However, in the case of small amounts of added water, the current density typically recovers in a few minutes to a steady state value and the anodization experiment finally leads to a highly-ordered structure (Fig. 4 – SEM images on the right side). These results can be explained by assuming that relatively small amounts of added water can rapidly evaporate from the hot $o$-H$_3$PO$_4$ electrolyte, so that the electrolyte reaches at the equilibrium a limited water content.

Typical $WO_3$ porous films (as those shown in Fig. 2-4) were characterized in view of their composition and structure. As-formed anodic films were in every case amorphous, as evident from XRD, HRTEM and SAED data in Fig. 5 (a-b). The conversion into crystalline structures could be obtained by a thermal treatment in air. Crystallization into monoclinic $WO_3$ could be achieved by annealing at 350°C (Fig. 5(a)).[17–19] For layers annealed at 450°C we observed significantly more intense reflections, ascribed to a higher crystallinity. The crystallization of these layers was further confirmed by clearly visible lattice planes of crystalline $WO_3$ in the HRTEM image (Fig. 5(c)). Also, the SAED pattern showed a four-fold symmetry with d spacing of 3.86, 2.70, 1.93 and 1.69 Å corresponding to the (002), (022), (004) and (042) planes of monoclinic $WO_3$.[20] Higher temperatures led to significant sintering and collapse of the nanochannels (Fig. S5-7).

XPS analysis of both as-formed and annealed layers confirmed the formation of $WO_3$ (Fig. 5(d)), with the $W4f_{7/2}$ and $W4f_{5/2}$ peaks centered at 36.71 and 38.91 eV (in line with the literature[21]), and a W:O ratio of *ca.* 1:3.4. Additionally, XPS analysis revealed the presence of P (*ca.* 1.1 at%) in both as-formed and annealed films (Fig. S8).[22] These results are well in line with those provided by EDX measurements, which revealed P contents of *ca.* 1.1-1.2 at% (Fig. S9) and are thus indicative of the formation of $WO_3$-$PO_4$ adsorbates at the interface (a $PO_4$-terminated surface of the oxide might also be the reason for the relatively large O/W atomic ratio compared to stoichiometric $WO_3$).

Overall, it can be concluded that the key role of the molten phosphate electrolyte is two-fold: *i)* it provides an environment with controlled and limited water content, and *ii)* it provides phosphate ions that protect the $WO_3$ layer from rapid dissolution. In fact we explored different phosphate sources and found that also other phosphorus-containing acids, namely, hot (nominally pure) pyro- and poly-phosphoric acids are also suitable electrolytes for the growth of such ordered porous $WO_3$ structures (Table S3 and Fig. S10).

Due to its electronic and optical properties, $WO_3$ has received large attention in the last decades in scientific and technological fields:[23] among other applications, it has been investigated as photocatalyst,[24] to fabricate electrodes for electrochromic devices[25,26] and for photo-electrochemical cells (*i.e.*, water splitting),[27] and it has been intensively studied also for its gas-sensing properties.[13,28–30] In these applications, such a $WO_3$ nanochannel structure, with its highly defined geometry, can represent a key for obtaining improved performance owing namely to advantageous directional charge transfer, enhanced gas diffusion and ion intercalating geometry, in comparison to devices fabricated from classical powder assemblies.

To illustrate a possible technological application, we fabricated from our $WO_3$ nanochannel layers $H_2$ gas-sensors (see the ESI). In comparison to previous literature reports on $WO_3$-based sensors (Pt- or Pd-decorated) showing reliable detection of $H_2$ only down to concentration of *ca.* 40 ppm at room temperature,[31] or down to 5 ppm at 250°C,[30] our Pt-contacted structures show nearly two-order (80-times) and one-order of magnitude lower limit of detection, respectively (see Fig. S14). Other applications may be especially in the field of catalysis and for fabricating photo-electrochemical devices (photo-anodes – Fig. S15).

In summary, in the present work we introduced a new anodization approach, based on the use of hot pure phosphoric acid as anodizing electrolytes, to form highly self-organized nanochannel structures of $WO_3$, and also of various other metal oxides including Al and Nb oxides. We showed that this effect is not limited to pure hot ortho-phosphoric acid but can be also achieved using pyro- and poly-phosphoric acids. The functionality and potential technological usefulness of the anodic porous films grown by this method were also briefly illustrated by fabricating highly sensitive $H_2$ gas-sensors.

ACKNOWLEDGMENT

Ulrike Marten-Jahns and Dr. Lei Wang are acknowledged for helping in the evaluation of the diffraction data. Helga Hildebrand and Dr. Anca Mazare are acknowledged for technical help and fruitful discussion on XPS data. Dr. Mirza Mackovic and Florian Niekiel are acknowledged for technical help in TEM analysis and data evaluation. Prof. Dr. Peter Wasserscheid (Institute of Chemical Reaction Engineering, Friedrich-Alexander University) and Marlene Scheuermeyer are acknowledged for technical help with Karl Fischer analysis. The authors would also like to acknowledge the ERC, the DFG and the Erlangen DFG cluster of excellence EAM for financial support. Dr. Marco Altomare and Prof. Dr. Elena Selli acknowledge financial support from MIUR through the 2009PASLSN PRIN project.

Electronic Supplementary Information

for

# Molten *o*-H$_3$PO$_4$ – A New Electrolyte for the Anodic Synthesis of Self-Organized Oxide Structures: WO$_3$ Nanochannel Layers and Others

M. Altomare,[§] O. Pfoch,[§] A. Tighineanu,[§] R. Kirchgeorg,[§] K. Lee,[§] E. Selli,[∥] and P. Schmuki*,[§,†]

[§]Department of Materials Science, Institute for Surface Science and Corrosion WW4-LKO, Friedrich-Alexander University, Martensstraße 7, D-91058 Erlangen, Germany

[†]Chemistry Department, Faculty of Sciences, King Abdulaziz University, 80203 Jeddah, Saudi Arabia Kingdom

[∥]Department of Chemistry, University of Milano, Via C. Golgi 19, I-20133 Milan, Italy

*Corresponding Author. Email: schmuki@ww.uni-erlangen.de

# C O N T E N T S

## Growth of anodic $WO_3$ layers on W foil

Preliminary anodization experiments were carried out on W foils in order to optimize the experimental conditions, that is, to obtain high aspect ratio $WO_3$ films with vertically aligned nanochannels. The results of preliminary anodization experiments are summarized in Tables S1-S3.

For these experiments, W foils (0.125 mm thick, 99.95% purity, Advent Research Materials LTD, Oxford, UK) were cut into 1 cm x 2 cm pieces, cleaned by ultra-sonication in acetone, ethanol and de-ionized water (10 min each) and finally dried in a $N_2$ stream.

For growing $WO_3$ nanochannels, the anodization experiments were carried out in a two-electrode electrochemical cell where the W film and Pt foil were the working and the counter electrodes, respectively. The two electrodes were immersed into the electrolyte in vertical configuration and placed at a distance of 2 cm from each other. The electrolyte was constantly stirred and kept at the desired temperature by thermostatic control provided by a heating-stirring plate. The heating plate was equipped with a thermocouple that was fully wrapped into a Teflon tape and immersed into the anodizing electrolyte. The experiments were performed under potentiostatic conditions, that is, by applying a constant direct current potential of 1-60 V (no sweeping) provided by a Volcraft VLP 2403 Pro power source. The resulting current density was recorded by using a Keithley 2100 6 ½ Digit multimeter interfaced with a laptop.

In most of the experiments, the electrolyte was composed of pure molten ortho-phosphoric acid (o-$H_3PO_4$, ≥ 99 %, Sigma-Aldrich). However, during preliminary anodization, a few additives were also added to the electrolyte. Precisely, DI $H_2O$ (18.2 MΩ cm), Ethylene Glycol (Fluka Analytical, ≥ 99.5 %) and Glycerol (Sigma Aldrich, ≥ 99.5 %) were used as additives for the *o-*$H_3PO_4$-based electrolytes. Also pyrophosphoric acid ($H_4P_2O_7$, ≥ 90 %, Fluka Analytical) and polyphosphoric acid (115 %, Sigma-Aldrich) were used (as received, *i.e.*, pure) as anodizing media (see Fig. S10).

**Growth of anodic WO$_3$ layers on non-conductive glass**

W layers (600-700 nm-thick, see Fig. S12) were deposited by e-beam evaporation (PLS 500 Labor System, Balzers-Peiffer, Germany) on non-conductive glass substrates at a pressure of 1-6 x $10^{-6}$ mbar and with a deposition rate of 0.1 nm s$^{-1}$. Tungsten granules (2-4 mm, 99.9 %, Chempur) were used as source of metal. Prior to the evaporation, the glass substrates (7.5 cm x 2.5 cm microscope glass slides, VWR) were cut into 1.2 cm x 2.5 cm pieces, cleaned (in acetone, isopropanol and DI H$_2$O for 10 min each) and finally dried in a N$_2$ stream.

The W films on non-conductive glass were all anodized in optimized conditions, that is, in pure molten *o*-H$_3$PO$_4$ at 5 V and 100°C, in a two-electrode electrochemical cell as reported above. The anodization experiments were run long enough to anodize through the entire thickness of the W layer, *i.e.,* not to have a metallic W layer left beneath the anodic film – this was obtained when the current density significantly dropped (Fig. S11). By anodizing for *ca*. 4 h in such experimental conditions, *ca*. 1.3 µm-thick transparent porous WO$_3$ layers were formed that exhibited excellent adhesion to the glass slide. These layers were used for fabricating the gas sensors. Such configuration of the device (see a sketch in Fig. S11) was adopted since preliminary experiments (WO$_3$ films on W foils or on FTO slides) showed the sensing measurements to be strongly affected by the presence of a conductive substrate. In particular, the response of devices fabricated on conductive substrates was relatively low, this because the current flowed preferentially through the conductive substrate, and changes in resistance of the anodic film upon exposure to analytes were relatively small.

**Characterization of anodic WO$_3$ layers**

A Hitachi field emission scanning electron microscope (FE-SEM S4800, Hitachi) was used for morphological characterization of the samples. The thickness of the anodic films was directly obtained from SEM cross-sectional micrographs. In the case of anodic WO$_3$ layers on W foils, the cross-sectional view was obtained by scratching off the anodic film with a blade. In the case of anodic WO$_3$ layers on non-conductive glass, the cross-sectional view was obtained either by scratching off the anodic film or by cracking the glass slides (for this, a diamond tip was used). Analysis of the morphological features of the WO$_3$ nanochannel structures (see Fig. 1-4 in main text) was performed by using the software Image J (data were then processed by Gaussian fitting). Energy-dispersive X-ray spectroscopy (EDAX Genesis, fitted to SEM chamber) was also used for the chemical analysis.

Transmission electron microscopy (TEM) was performed by using a Philips CM300 UltraTWIN, equipped with a LaB6 filament and operated at 300 kV. TEM images and selective area diffraction (SAED) patterns were recorded with a fast scan (type F214) charge-coupled device camera from TVIPS (Tietz Video and Image Processing Systems), with an image size of 2048 x 2048 pixels. For TEM investigations, the samples were mechanically scratched from the substrate and the resulting powder was deposited on copper TEM grids coated with lacey carbon film.

X-ray diffraction (XRD) patterns were collected using an X'pert Philips PMD diffractometer with a Panalytical X'celerator detector, using graphite-monochromized Cu K$\alpha$ radiation ($\lambda$ = 1.54056 Å).

Composition and chemical state of samples were determined by X-ray photoelectron spectroscopy (XPS) using a PHI 5600 Multi-Technique System (Physical Electronics, USA) equipped with a monochromatic Al K$\alpha$ X-ray source (1486.6 eV).

**Fabrication of gas-sensors**

As shown in Fig. S11, the fabrication of the gas-sensing devices was completed by annealing the porous $WO_3$ layers and by depositing Pt electrodes on their top.

XRD analysis showed that as-formed anodic films are typically amorphous (Fig. 5(a-b) in main text) and the gas-sensing experiments revealed that, along with Pt deposition onto the semiconductor (see below), the crystallization of the oxide is a key to fabricate functional gas-sensors (Fig. S14). Thus, the $WO_3$ films were annealed in air at different temperatures (250-650°C range) for 1 h, with a heating/cooling rate of 30 °C min$^{-1}$ by using a rapid thermal annealer (RTA, Jipelec JetFirst100).

Then, the anodic films were contacted on top by depositing through a mask two 200 nm-thick Pt electrodes by using a Leica EM SCD 500 plasma sputtering system (Fig. S13). The deposition was carried out at 16 mA and at a rate of 0.16-0.20 nm s$^{-1}$ (vacuum conditions, 10$^{-2}$ mbar of Ar).

**Gas-sensing experiments**

For the gas-sensing measurements, a device was placed into the sensing chamber (Scheme S1) and flushing with artificial air (Linde, Germany) was carried out until reaching a constant resistance measured by the device. The chamber was kept under thermostatic control by using a Eurotherm 3216-based temperature controller (Invensys Eurotherm, USA).

Gold wires were connected to the Pt electrodes that were sputter-deposited onto the annealed anodic $WO_3$ films. The resistance of the Pt/anodic film system was measured with a Keithley 2400 Source Meter (Keithley Instruments, USA) by applying a bias of 1 V. Preliminary experiments showed that the I-V characteristics of the $WO_3$ layers exhibited an ohmic response in this range of applied bias (not shown).

To investigate the device sensitivity, the Pt/WO$_3$ films were exposed to different concentration of H$_2$. For this purpose, different amounts of H$_2$-Ar mixtures (90.000 ppm of H$_2$ in Ar, Linde, Germany) were injected into the background stream of artificial air. The flow of the different gases was controlled with digital mass flow controllers MF1 (MKS Instruments, Germany). The relative response "$r$" and sensitivity "$s$" of the sensors were calculated according to following equation:

$$r = \frac{R_0 - R}{R_0} \cdot 100 \ \% \qquad \text{Eq. (1)}$$

$$s = \frac{\delta(r)}{\delta(H_2 \text{conc.})} \qquad \text{Eq. (2)}$$

where "$R_0$" and "$R$" are the resistance of the sensor when exposed to background air flow and to the H$_2$ injection, respectively.

## Table S1 – Preliminary experiments in pure $o$-H$_3$PO$_4$

Anodization experiments carried out on W foils in pure molten $o$-H$_3$PO$_4$[§].

| T (°C) | E (V) | TIME (h) | DESCRIPTION | THICKNESS |
|---|---|---|---|---|
| 20 | 20 | 0.5 | Little extent of etching | Not measured |
| | 30 | | | |
| | 40 | | Large extent of etching | |
| 20 | 5 | 1 | Compact oxide layer | Few tens of nm |
| 40 | | | | |
| 60 | | | | |
| 80 | | | Porous layer | Few tens of nm |
| 80 | 20 | 1 | Channels with a little extent of etching at the top | 400 – 500 nm |
| | 40 | | Large extent of etching | Not measured |
| 100 | 1 | 1 | Porous layer | Few tens of nm |
| | 2.5 | | Highly ordered nanochannels, smooth and flat at the top | 300 nm |
| | 5 | | | 400 – 600 nm |
| | 5* | | | |
| | 10 | | | 0.6 – 1.0 µm |
| | 10 | 4 | | 0.8 – 1.2 µm |
| | 15 | 1 | Nanochannels with a little extent of etching at the top | 1.1 – 1.3 µm |
| | 20 | | Porous layer | 1.2 – 1.4 µm |
| 100 | 5 | 0.5 | Porous layer | 150 – 200 nm |
| | | 2 | Highly ordered nanochannels, smooth and flat at the top | 600 – 700 nm |
| | | 4 | | 1.3 – 1.5 µm |
| | | 8 | | 1.9 – 2.1 µm |
| | | 12 | Ordered nanochannels with nanograss[†] at the top | 2.1 – 2.6 µm |
| | | 24 | | 3.4 – 3.6 µm |
| | 2.5 | 24 | | 1.0 – 1.1 µm |
| 120 | 5 | 1 | Nanochannels with a little extent of etching at the top | 400 – 500 nm |
| 90 | 5 | 1 | Highly ordered nanochannels, smooth and flat at the top | 100 – 200 nm |
| 95 | | | | 200 nm |
| 105 | | | | 400 – 450 nm |
| 110 | | | | 1.0 – 1.1 µm |
| 110[‡] | | | | 800 – 900 nm |
| 115 | | | Ordered nanochannels with grass at the top | 1.0 – 1.3 µm |

[§]$o$-H$_3$PO$_4$ = ortho-phosphoric acid; solid at room temperature (melting point = 42.3°C).
*the electrolyte was aged by anodizing a W foil for 100 h at 2.5 V and 100°C.
[‡]the electrolyte was aged by anodizing a W foil for 25 h at 5V and 110°C.
[†]the term "nanograss" refers to needle-like nanosized structures that form at the top of the anodic layer as consequence of considerable etching (typically observed for extended anodization experiments).

**Table S2 – Preliminary experiments in *o*-H$_3$PO$_4$–based electrolytes**

Anodization experiments carried out on W foils in *o*-H$_3$PO$_4$-based electrolytes.

| ADDITIVE | T (°C) | E (V) | TIME (h) | DESCRIPTION | THICKNESS |
|---|---|---|---|---|---|
| H$_2$O traces* | 110 | 5 | 1 | Ordered nanochannels with grass at the top | 1.0 – 1.1 µm |
| 0.5vol.%H$_2$O§ | 100 | | | Highly ordered nanochannels, smooth and flat at the top | 300 – 400 nm |
| 1vol.%H$_2$O§ | 20 | 5 | 0.5 | Compact oxide layer | Not measured |
| | | 20 | | | |
| | | 40 | | Little extent of etching | |
| 2vol.%H$_2$O§ | 100 | 5 | 1 | Highly ordered nanochannels, smooth and flat at the top | 500 – 600 nm |
| | | 10 | | | |
| | | 15 | | Channels with a little extent of etching at the top | |
| 5vol.%H$_2$O§ | 20 | 5 | 0.5 | Compact oxide layer | Not measured |
| | | 20 | | | |
| | | 40 | | Little extent of etching | |
| | | 60 | | Electropolishing | |
| 10vol.%H$_2$O§ | 20 | 5 | 0.5 | Compact oxide layer | Not measured |
| | | 20 | | | |
| | | 40 | | Large extent of etching | |
| | 100 | 5 | 1 | Ordered nanochannels with grass at the top | 0.9 – 1.1 µm |
| | | 10 | 0.5 | | 0.6 – 1.1 µm |
| 25vol.%EG‡ | 100 | 5 | 1 | Highly ordered nanochannels, smooth and flat at the top | 300 – 400 nm |
| | | 20 | 0.3 | Large extent of etching | Not measured |
| 25vol.%GLY† | 50 | 10 | 1 | Compact oxide layer | Not measured |
| | | 20 | | | |
| | 80 | | | Porous layer | Few tens of nm |
| | 100 | 10 | 0.3 | Ordered nanochannels, smooth and flat at the top | 300 – 400 nm |

*the electrolyte was exposed overnight to ambient air to allow for moisture uptake and no preliminary heat treatment was performed before anodizing.
§the water content (in terms of vol.%) represents a nominal value, calculated by taking into account the volume of pure water added to the electrolyte, and by assuming the latter to be water-free (see Fig. S4 for more details on the electrolyte water content).
‡EG = Ethylene glycol.
†GLY = Glycerol.

**Table S3 – Preliminary experiments in other P-containing acids**

Anodization experiments carried out on W foils in other hot pure phosphorus-containing acid electrolytes.

| ACID | T (°C) | E (V) | TIME (h) | DESCRIPTION | THICKNESS |
|---|---|---|---|---|---|
| $H_4P_2O_7$* | 100 | 5 | 1 | Porous layer | Few tens of nm |
| | | 20 | | Channels with a large extent of etching | 250 – 300 nm |
| | | 40 | 0.3 | Large extent of etching | Not measured |
| | 110 | 5 | 1 | Highly ordered nanochannels, smooth and flat at the top | 1.0 – 1.1 µm |
| | | 10 | | | |
| | | 20 | | Channels with a large extent of etching | 450 – 550 nm |
| | 120 | 5 | | Highly ordered nanochannels, smooth and flat at the top | 1.1 – 1.2 µm |
| POLY[‡] | 120 | 5 | 1 | Ordered nanochannels, smooth and flat at the top | 200 – 250 nm |

*$H_4P_2O_7$ = pyrophosphoric acid; solid at room temperature (melting point = 71.5°C).
[‡]POLY = polyphosphoric acid; viscous liquid at room temperature.

**Fig. S1 – TEM analysis**

TEM images of a WO$_3$ nanochannel layer formed by anodization of W in molten pure *o*-H$_3$PO$_4$ at 5 V and 100°C.

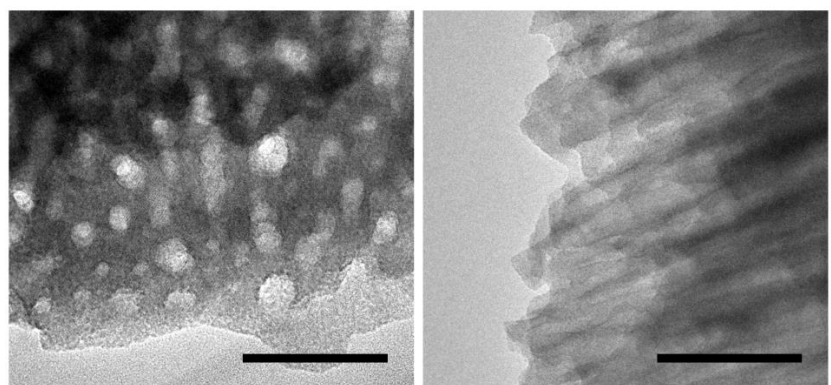

(scale bars = 50 nm)

**Fig. S2 – Effect of electrolyte temperature**

SEM images of WO$_3$ films grown by anodization of W foils in pure molten *o*-H$_3$PO$_4$ at 5 V for 1 h, at different temperatures of the electrolyte (60-120°C). In the case of the anodic layer formed at 80°C, no clear cross sectional image could be taken (*i.e.*, a few nm-thick film).

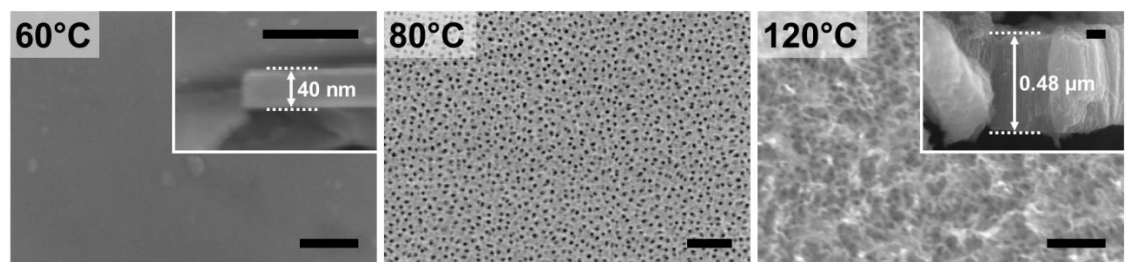

(scale bars = 100 nm)

## Fig. S3 – J-time profiles

Current density (J) *vs.* time profiles of anodization experiments carried out on W foils in molten pure *o*-$H_3PO_4$ at 100°C for 1 h, at different applied potential in the 1-20 V range (formed structures are shown in Fig. 2 in the main text).

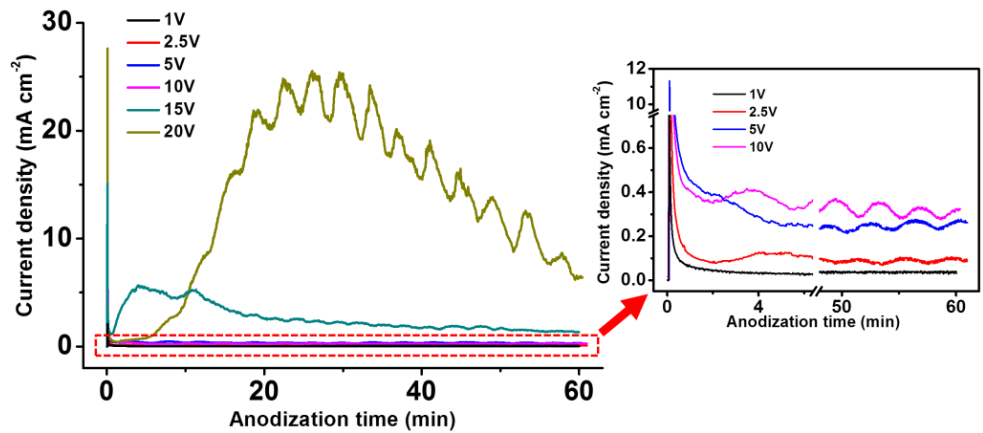

The oscillation of the current density (visible also in the main text in Fig. 4, and in Fig. S11) is only an artifact given by the temperature feed-back control of the thermocouple. The latter is interfaced to the heating-stirring plate used for the anodization experiments, and allows for maintaining the electrolyte temperature at the desired value. However, slight control drifts take place, and the temperature of the molten *o*-$H_3PO_4$ typically oscillates by ± 3°C (this may be also ascribed to the relatively high viscosity of the electrolyte which limits heat transfer). The current density measured during the anodization experiments mirrors such oscillating trend, being J strictly related to the electrolyte temperature.

**Fig. S4 – Karl Fischer analysis (water content determination)**

Water content (in ppm and %) of nominally pure $o$-$H_3PO_4$, molten and heated up to 60 and 100°C, determined by using a 756 KF Coloumeter (Metrohm).

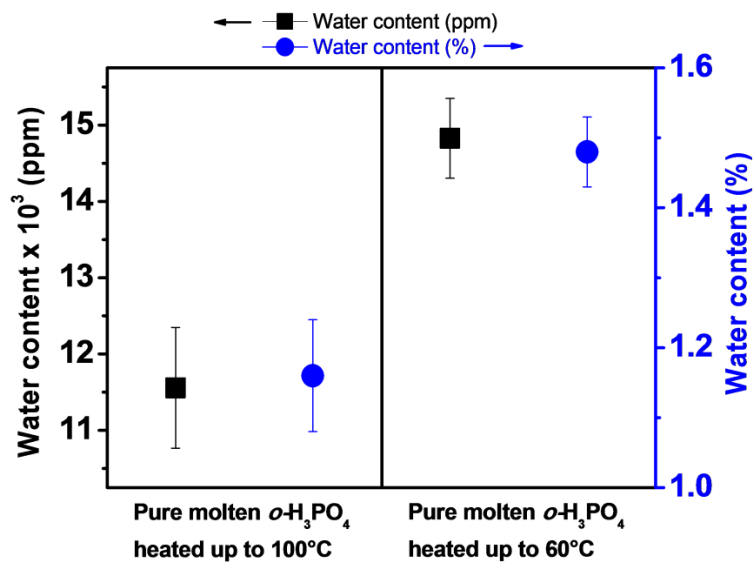

**Fig. S5 – XRD analysis – WO$_3$ films on glass slides**

XRD patterns in the 10-30 degree region of WO$_3$ nanochannel layers grown by anodization of W in pure molten $o$-$H_3PO_4$ at 5 V and 100°C, and annealed at different temperatures in the 250-650°C range (air, 1 h).

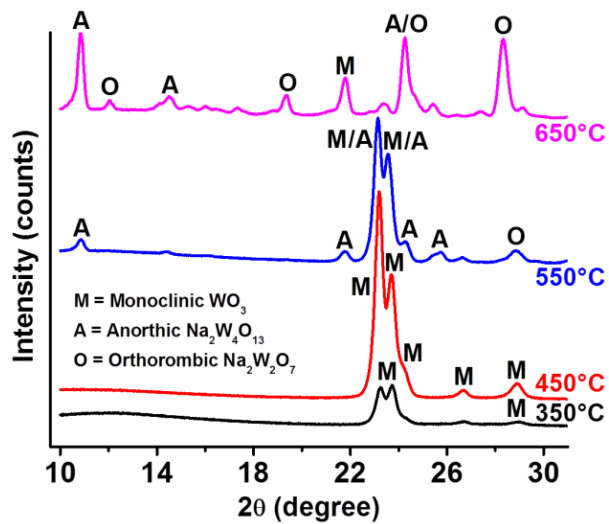

WO$_3$ layers grown on glass slides were shown to convert also into various sodium tungstate species when crystallized at relatively high temperature (*e.g.*, T ≥ 550°C).

## Fig. S6 – XRD analysis – WO$_3$ films on W foils and on quartz slides

XRD patterns of as-formed and annealed WO$_3$ nanochannel layers grown by anodizing W foils and W films e-beam evaporated onto quartz slides. The anodization experiments were carried out in pure molten *o*-H$_3$PO$_4$ at 5 V and 100°C. The WO$_3$ layers were annealed at different temperatures in the 450-650°C range (air, 1 h). The inset shows the XRD patterns in the 10-22 degree region.

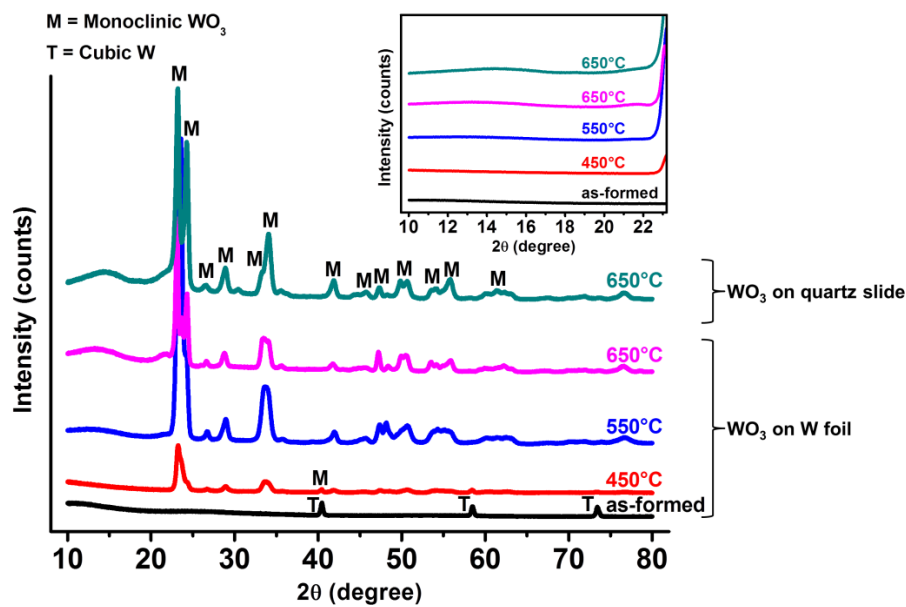

Conversely to WO$_3$ films grown on glass slides, the WO$_3$ nanochannel layers grown on W foils or on quartz slides were shown to convert into monoclinic phase only when annealed at T ≥ 450°C, and no formation of sodium tungstate species could be observed (compare the XRD patterns in the 10-22 degree region to data in Fig. S5, and see Fig. S8 for more discussion).

**Fig. S7 – Effect of annealing temperature**

SEM images of WO$_3$ nanochannel layers anodically grown on non-conductive glass substrate and annealed at different temperatures (450-650°C, in air, 1 h).

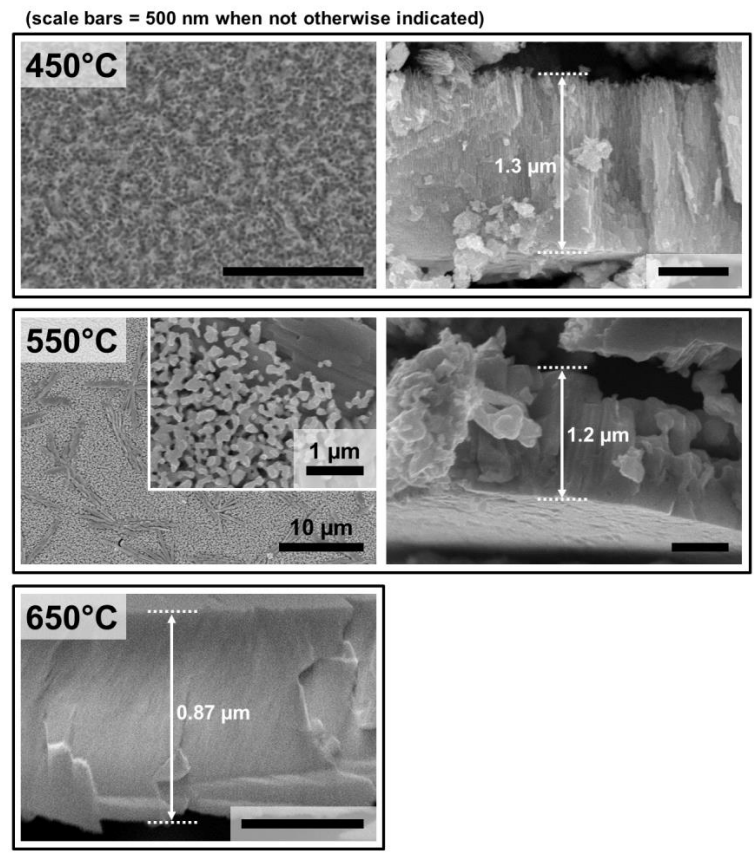

The nanochannel structures were shown to withstand thermal treatment at temperatures up to 450°C. On the contrary, high annealing temperatures (≥550°C) led to significant sintering and collapse of the structures.

Besides the effect on the morphological features, relatively high annealing temperature (≥550°C) led also to formation of sodium tungstate species (these with different stoichiometry and crystallographic features) as proved by appearance in the XRD patterns of intense reflections mainly at 10.8, 24.3 and 28.3 degree (see Fig. S5).[1] These results can be explained by assuming that at such annealing temperatures, solid state reaction occurs that leads to diffusion of Na ions from the glass substrate through the oxide film. This is in agreement with $WO_3$ crystallographic features and consequent ability to intercalate $Na^+$ ions.[2]

**Fig. S8 – XPS analysis**

High resolution XPS spectra in the W4f, O1s and P2p regions of $WO_3$ nanochannel layers (as-formed and annealed at 450°C, air, 1 h) grown by anodization of W in pure molten *o*-$H_3PO_4$, at 5 V, 100°C.

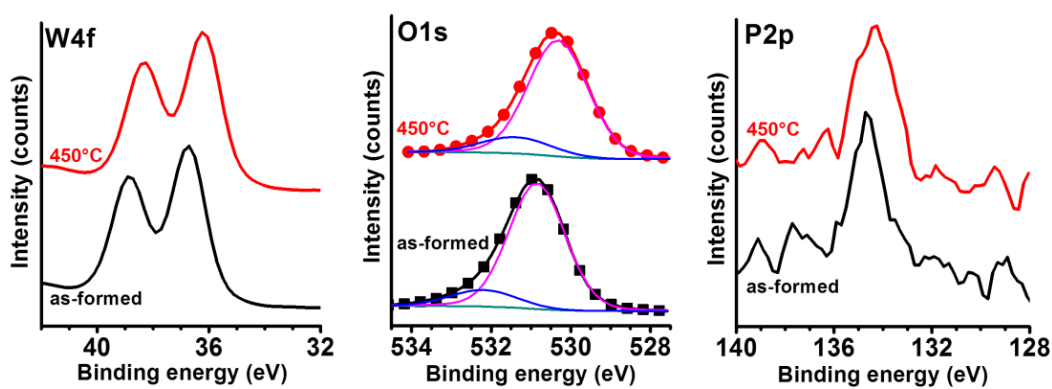

The survey spectrum of the film annealed at 450°C (Fig. 5(d) in main text) shows Na1s signal. As mentioned above (Fig. S5 and S7), one can explain this assuming that the thermal treatment at 450°C enables $Na^+$ intercalation. On the other hand, the film annealed at this temperature was shown to be composed of monoclinic $WO_3$ (Fig. 5(a,c)). Thus we assume that only higher annealing temperatures (≥550°C) enable intercalation of significant amounts of $Na^+$ ions with consequent transformation of $WO_3$ into stoichiometric $Na_2W_2O_7$.

The high resolution XPS spectra show that after annealing at 450°C a *ca*. 0.5 eV-shift of W4f doublet towards lower binding energies can be seen. According to the literature, a shift toward higher binding energies is typically observed after annealing, that is ascribed to complete oxidation of the anodic film to form stoichiometric $WO_3$ and also to a reduction of hydroxide species and water adsorbed at the oxide surface.[3] Therefore, the shift towards lower binding energies can be due to intercalation of Na ions that induces partial reduction of $WO_3$ (in line with XRD and XPS data).

The O1s signals were deconvoluted into two peaks by Gaussian fitting: the main peaks at 530.88 and 530.28 eV are attributed to lattice oxygen and are in line with the literature on anodic $WO_3$.[4,5] On the other hand, the peaks at higher binding energies, *i.e.*, at *ca.* 532.27-531.48 eV, are attributable to adsorbed hydroxide.[4,5]

**Fig. S9 – EDX analysis**

EDX spectra and relative data for as-formed and annealed (450°C) $WO_3$ nanochannel layers that were anodically grown on W foils in pure molten *o*-$H_3PO_4$, at 5 V, 100°C. The peaks of Aluminum are ascribed to the aluminum sample holder.

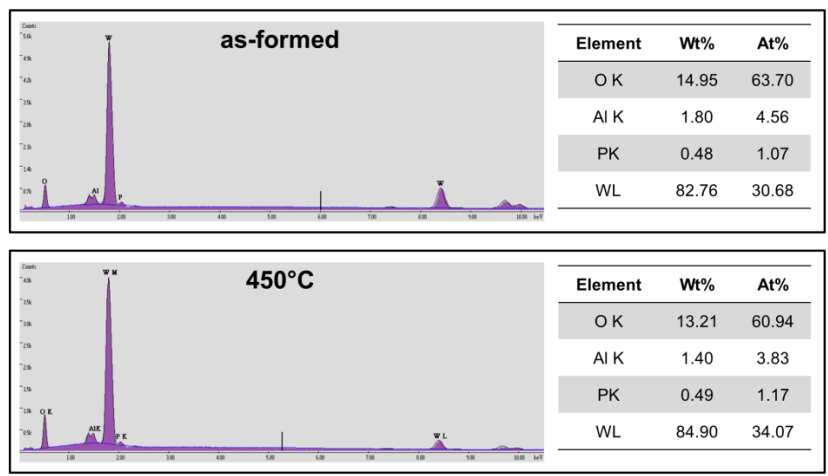

**Fig. S10 – Anodization in pyro- and poly-phosphoric acid**

SEM images of WO$_3$ nanochannel layers grown by anodizing W foils at 120°C (5 V, 1 h) in (nominally) pure pyro- and poly-phosphoric acid.

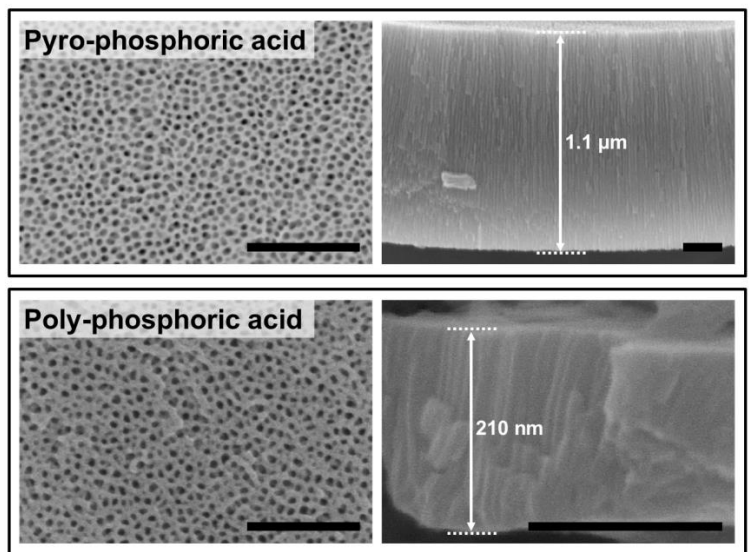

(scale bars = 200 nm)

**Fig. S11 – Gas-sensor fabrication**

(a) Sketch of the different steps for fabricating the gas-sensors; (b) J-time profile recorded during the fabrication of $WO_3$ nanochannel layer on a non-conductive glass slide and optical pictures of the obtained porous transparent film (c) Sketch of the resistive gas-sensor configuration adopted in this work, that is, with the anodic nanoporous $WO_3$ film grown on non-conductive glass, compared to an alternative configuration which implies the direct growth of the anodic layer on the W metal foil or, more generally, on a conductive substrate.

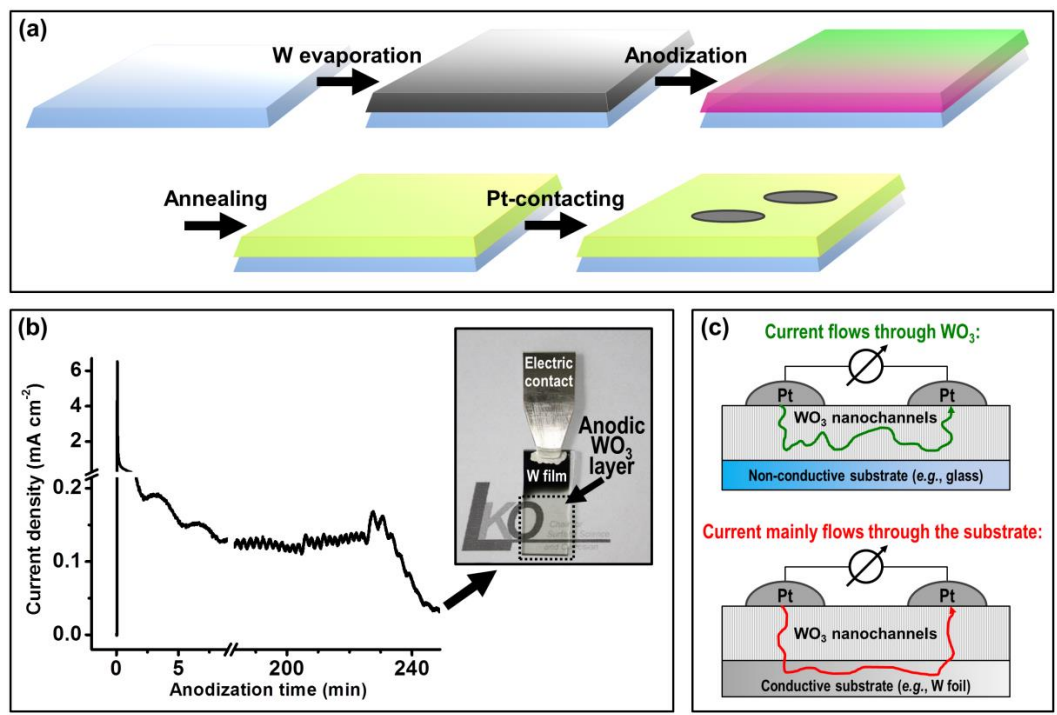

## Fig. S12 – W and WO$_3$ films on glass

Top view and cross sectional SEM images of (a) 600 nm-thick W film evaporated onto non-conductive glass slide, and (b) 1.3 µm-thick WO$_3$ nanochannel layer formed by complete anodization (*ca.* 4 h) of the W film in pure molten *o*-H$_3$PO$_4$ at 5 V and 100°C. The volume expansion is ascribed to the lower density of the metal oxide compared to that of the metal.[6]

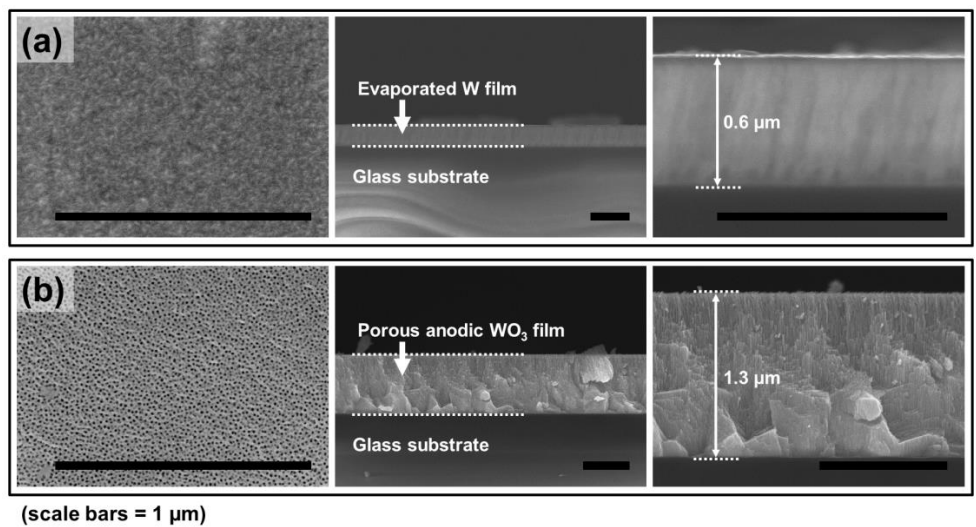

(scale bars = 1 µm)

## Fig. S13 – Gas-sensing device

(a) Optical and (b,c) top view SEM pictures of a gas-sensing device fabricated from WO$_3$ nanochannel layers anodically grown in pure molten *o*-H$_3$PO$_4$ (5 V, 100°C) on non-conductive glass substrates; (b) Low magnification SEM image of the device showing the anodic layer and the two sputtered Pt electrodes; (c) High magnification SEM image of the sputtered Pt electrode.

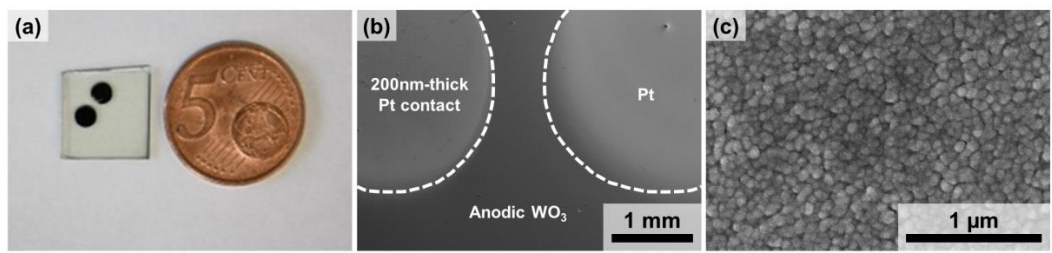

**Scheme S1 – Gas-sensing setup**

Sketch of the gas-sensing setup.

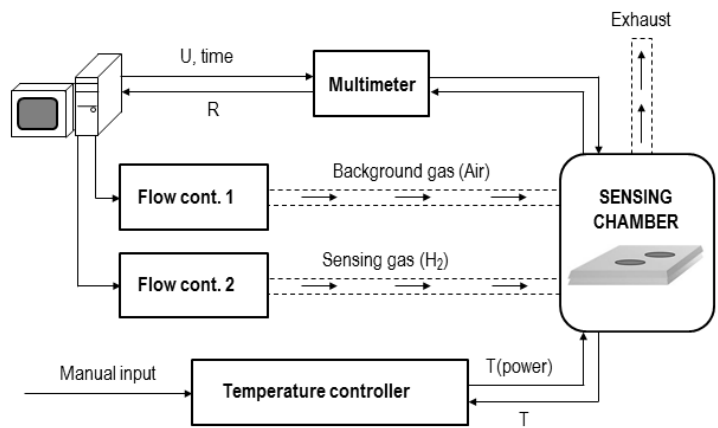

**Fig. S14 – Gas-sensing results**

$H_2$ sensing response of layers annealed at different temperatures (250-650°C) to $H_2$ injections of 2.3 ppm, at a sensing temperature of 120°C. Up-right: sketch of the gas-sensor; Inset: response of a sensor crystallized at 450°C to $H_2$ injections of different concentrations (500 ppb - 2.3 ppm) at a sensing temperature of 120°C.

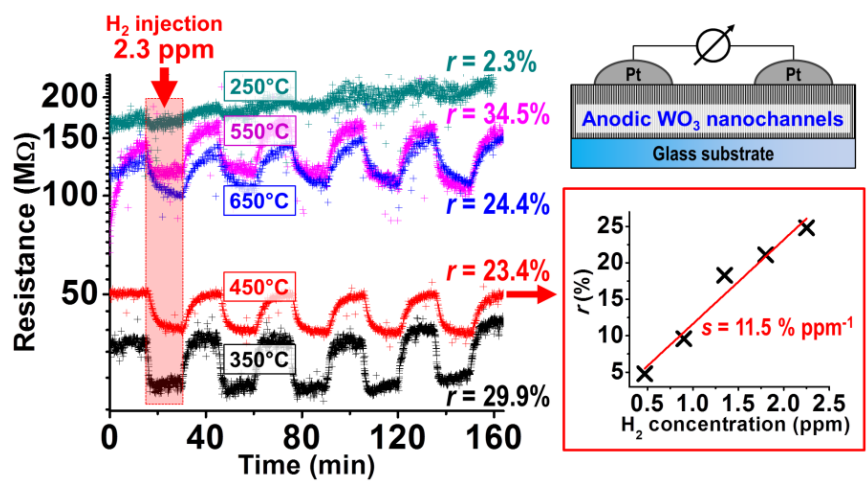

Overall, the sensing results were well in line with the $H_2$ sensing mechanism reported in the literature.[7–9] Briefly, the sensing ability of $WO_3$ is generally ascribed to reduction of sensor resistance upon exposure to $H_2$ (reducing gas). This occurs because $H_2$ is oxidized by the oxygen species adsorbed at the surface of the oxide, forming water as final product ("receptor" function).[7] Thus, the consumption of adsorbed oxygen leads to the increase of conductivity of the sensor ("transducer" function). A similar effect can be brought about by most of the acidic gases (*e.g.*, $H_2S$, $NO_x$, *etc.*). Also, Pt is typically deposited onto the $WO_3$ structures, this with a two-fold effect: *i)* it improves the sensitivity and reduces the response time of the sensors; and *ii)* it increases the selectivity of the device towards $H_2$, this because $H_2$ adsorption and consequent oxidation is particularly triggered by Pt deposits.[10] More precisely, the increase of sensitivity occurs not only through a direct electronic interaction between Pt and the semiconductor surface (enabling a strong catalytic Schottky contact), but also by "spill-over effect", that means that Pt remarkably facilitates the oxidation of $H_2$. This typically leads to injection, into the $WO_3$ conduction band, of larger amount of electrons upon exposure to $H_2$, and therefore to an enhanced response of the sensing device.[10]

A series of preliminary experiments showed the sensors to deliver a reliable signal at relatively low sensing temperature: stable and reproducible response as high as of *ca.* 23% to $H_2$ injections of 2.3 ppm was measured at 80-120°C.

Concerning the effect of thermal treatment on the sensing ability (Fig. 4 in main text), devices annealed at relatively high temperatures (≥550°C) delivered a slow drop and recovery of the resistance while films treated at 250°C showed poor response. These results correspond well with XRD and SEM analysis, that is, films annealed at temperature ≥550°C underwent sintering (significant loss of porosity) and conversion into tungstate species (Fig. S5, S7 and S8), while structures treated at 250°C resulted not crystalline (Fig. 5(a) in main text).

Instead, a proper annealing treatment at 350-450°C, led the WO$_3$ channels to deliver the best response, in terms of both speed and magnitude, that is, a large and markedly fast resistance-drop and recovery upon exposure to H$_2$ pulses of a concentration as low as 2.3 ppm.

The gas-sensing ability of devices crystallized at 450°C was also assessed upon exposure to H$_2$ concentration in the 0.5-2.3 ppm range (Inset): the response to H$_2$ pulses of 500 ppb was of *ca.* 5% and the devices also showed, in this range of H$_2$ concentration, a linear correlation between the response and the analyte amount (sensitivity $s = 11.5$ % ppm$^{-1}$).

In comparison to previous literature reports on WO$_3$-based sensors (Pt- or Pd-decorated) showing reliable detection of H$_2$ only down to concentration of *ca*. 40 ppm at room temperature,[11] or down to 5 ppm at 250°C,[12] our Pt-contacted structures show nearly two-order (80-times) and one-order of magnitude lower limit of detection, respectively.

**Fig. S15 – PEC water splitting**

Photo-electrochemical (PEC) water splitting results measured with WO$_3$ nanochannel layers grown by anodization of W foils in pure molten *o*-H$_3$PO$_4$ at 5 V and 100°C. The anodization experiments lasted 4 h and the layers thickness was of *ca.* 1-1.3 µm. Prior to the PEC measurements, the WO$_3$ layers were annealed at 450°C (air, 1 h) to convert the anodic structures into a monoclinic WO$_3$ phase.

The PEC water-splitting ability of the crystalline anodic films was investigated in aqueous 0.5 M Na$_2$SO$_4$ (containing also 0.1 M HCOONa), with a three-electrode configuration, consisting of a WO$_3$ photo-anode used as working electrode, a saturated Ag/AgCl electrode as reference and a platinum foil as counter electrode. An external bias, provided by a scanning potentiostat (Jaissle IMP 88 PC) with a scan rate of 1 mV s$^{-1}$, was applied to the PEC cell to record the photocurrent response (and also the dark current). The photocurrent density was calculated by dividing the photocurrent response (in mA) by the anodic oxide area (0.385 cm$^{-2}$) irradiated by the light source. The experiments were carried out under simulated AM 1.5 illumination provided by a solar simulator (300 W Xe lamp with a Solarlight optical filter) with an irradiation power of 100 mW cm$^{-2}$. The light intensity was measured prior to the experiments using a calibrated Si photodiode.

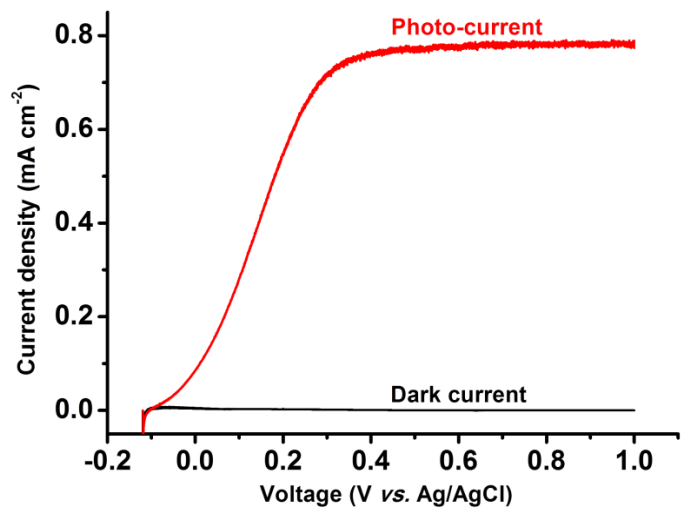

The crystalline *ca*. 1-1.3 μm-thick WO$_3$ nanochannel layers deliver a positive photo-current under solar light irradiation, and under anodic bias (relative to the flat-band potential). These results confirmed the n-type behavior of the anodic WO$_3$ structures.[13]